\def\swift{{\it Swift}}
\def\xmm{{\it XMM-Newton}}

\def\keV{{\rm \ keV}}
\def\1h{{1H0707-495}}
\def\et{{et al.\ }}
\def\mcg{{MCG--6-30-15}}
\def\mrk335{{Mrk~335}}

\def\feka{{Fe~K$\alpha$}}

\documentclass[useAMS,usenatbib]{mn2e}
\usepackage{graphicx}

\title[X-ray/UV Correlation in 1H0707-495]{Searching for Correlations in Simultaneous X-ray and UV Emission in the Narrow-line Seyfert 1 galaxy 1H~0707--495}
\author[D. Robertson et al.]{D.R.S. Robertson$^{1,2}$\thanks{E-mail:
roberd9@mcmaster.ca}, L.C. Gallo$^{1}$, A. Zoghbi$^{3}$ and A.C. Fabian$^{4}$\\
$^{1}$Department of Astronomy \& Physics, Saint Mary's University, 923 Robie Street, Halifax, NS B3H 3C3, Canada\\
$^{2}$Department of Physics \& Astronomy, McMaster University, Hamilton, ON L8S 4M1, Canada\\
$^{3}$Department of Astronomy, University of Michigan, 1085 South University Avenue, Ann Arbor, MI 48109, USA\\
$^{4}$Institute of Astronomy, University of Cambridge, Madingley Road, Cambridge CB3 0HA}

\begin{document}

\date{DRAFT}

\pagerange{\pageref{firstpage}--\pageref{lastpage}} \pubyear{2002}

\maketitle

\label{firstpage}

\begin{abstract}
We examine simultaneous X-ray and UV light curves from multi-epoch 8 day \xmm\ observations of the narrow line Seyfert 1 galaxy 1H~0707--495. The simultaneous observations reveal that both X-ray and UV emission are variable and that the amplitude of the X-ray variations is significantly greater than that of the UV variations in both epochs. Using a discrete correlation function (DCF) the X-ray and UV light curves were examined for correlation on timescales up to 7.0 d. Low significance ($\sim 95$ per cent confidence) correlations with the UV leading the X-ray variations are observed.  The lack of a significant correlation between the UV and X-ray bands seems consistent with the X-ray source being centrally compact and dominated by light bending close to the black hole.  In addition, multi--band X-ray light curves were examined for correlations on similar timescales.   Highly significant ($> 99.9$ per cent confidence) correlations were observed at zero lag consistent with previous studies of this AGN.

\end{abstract}


\section{Introduction}

Active galactic nuclei (AGNs) are characterised by their broad spectral energy distribution that extends over all observable wavelengths.  In the standard paradigm the UV emission is attributed to thermal emission from the accretion disc, while the X-rays originate from Comptonisation of UV disc photons in a centrally compact corona of relativistic electrons.  While the emitting regions may be physically distinct interplay between them is predicted.  If the corona indeed creates X-ray emission through Comptonisation of disc photons then one would expect that variations seen in the UV emission would propagate into the X-ray producing corona after some time delay.  Similarly, some X-ray photons emitted by the corona will inevitably illuminate the accretion disc where they will be reprocessed and seen in the UV.  Variability in X-ray will then be echoed in UV at a later time corresponding to light travel time between the regions if the reprocessing timescales are negligible.

\begin{figure*}
\includegraphics[width=1.0\textwidth]{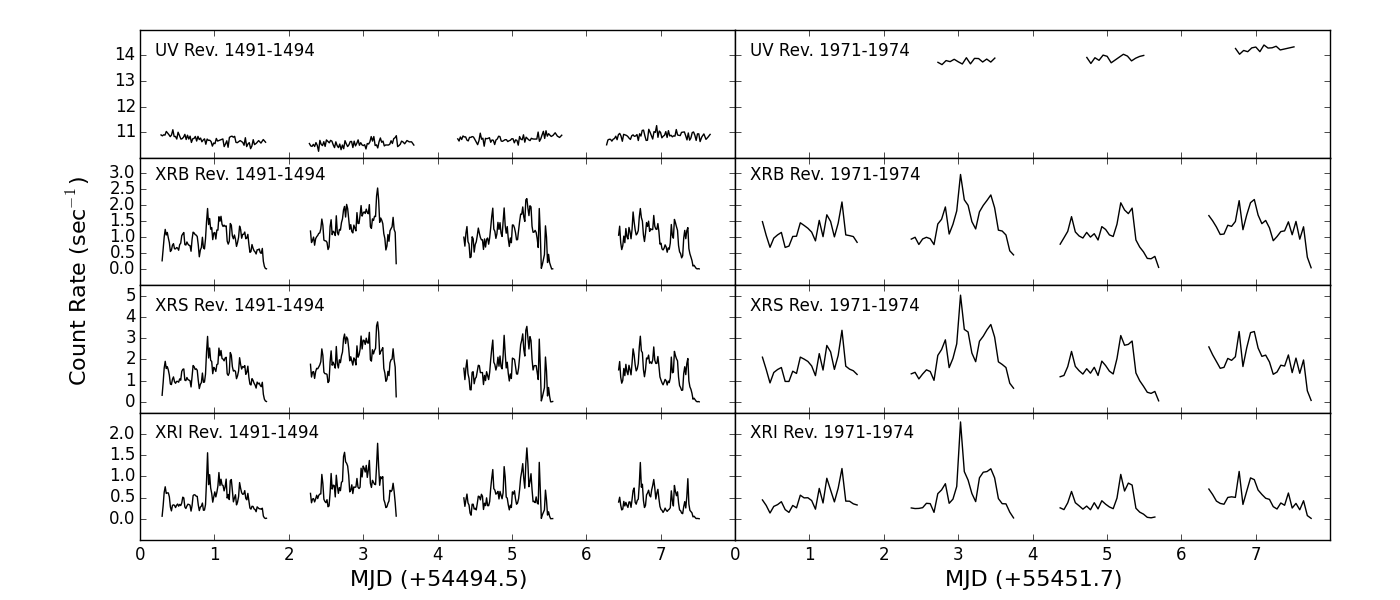}
\caption{1H~0707--495 UV (UVW1) and X-ray light curves at both epochs, Rev. 1491--1494 (left panel) and Rev. 1971--1974 (right panel). Errors in count rate are too small to be seen.}
\label{fig:lc}
\end{figure*}

\begin{table*}
\caption{\xmm\ observation log for 1H~0707--495. The * denotes X-ray observations affected by high radiation.}
\begin{center}
\begin{tabular}{c|c|c|c|c|c|c}
Rev & Observation ID& Start Date (UTC) & Duration (s) & UV Filter&UV Exposures&UV Exposure Length (s)\\
\hline
1491&0511580101&2008/01/29 18:10:59&121591&UVW1&72&1200\\
1492*&0511580201&2008/01/31 18:03:18&105713&UVW1&75&1200\\
1493*&0511580301&2008/02/02 17:55:35&104157&UVW1&75&1200\\
1494*&0511580401&2008/02/04 17:57:46&101785&UVW1&75&1200\\
1971*&0653510301&2010/09/13 00:14:45&128467&UVW1&0&-\\
1972*&0653510401&2010/09/15 00:16:54&127867&UVW1&15&4400\\
1973*&0653510501&2010/09/17 00:19:15&127268&UVW1&13&4400\\
1974*&0653510601&2010/09/19 00:18:10&126867&UVW1&13&4400\\
\hline
\end{tabular}
\end{center}
\label{tab:obslog}
\end{table*}%

The search for correlation between UV and X-ray emission has generated mixed results. Smith \& Vaughan (2007; hereafter SV07) examined a sample of ten Seyfert 1 galaxies with simultaneous X-ray and UV data obtained with \xmm\ using a discrete correlation function (DCF; Edelson \et 1988).  Their work was mainly sensitive to short time scale correlations, less than $\sim 2$ days.  SV07 found only one object, NGC~4051, that exhibited a modestly significant lag ($\sim68 - 95$ per cent confidence), with the X-ray leading the UV by about 0.2~d.  These findings corroborated earlier conclusions by Mason \et (2002), which found a similar 0.2~d lag in NGC~4051 with $\sim85$ per cent confidence.  On the other hand, SV07 refute a 1.6~d lag (UV leading X-ray) in \mcg\ that was earlier reported by Ar\`evalo \et (2005).

Further work on NGC~4051 utilising fifteen \xmm\ observations over 45 days revealed the UV lagging X-ray by $\sim3$~ks (Alston \et 2013). Alston \et find the lag is consistent with thermal reprocessing of X-rays. Breedt \et (2010) examined NGC~4051 on longer time scales using twelve years of monitoring data and find two significant correlated peaks at $\sim1.2$ and $39$~d.  In both cases the X-ray lead the UV variations also suggesting X-ray reprocessing in the disc and torus. However, Alston \et did not find any significant correlations on similar timescales.

Cameron \et (2012) examined one of the smallest known supermassive black holes, NGC~4395, using \swift. Over the course of a year from April 2008 to March 2009 Cameron \et collected data in the UVW2 filter (212~nm rest frame) and X-ray (2-10 keV) to probe long term variability between the wavebands. Using a cross correlation function Cameron \et were able to determine a correlation of zero lag between the UV and X-ray regimes, only when re-binning did they find a lag of 0.005~d with 99 per cent confidence over $\pm40$ days.

Other works examining various time scales have found evidence for UV leading X-ray (e.g. Marshall \et 2008; Shemmer \et 2003); X-ray leading UV (Nandra \et 1998; Shemmer \et 2001, Breedt \et 2010; Lohfink \et 2014); and significantly correlated light curves with zero lags (e.g. Breedt \et 2008). While in theory an examination for correlations would appear straightforward, in practise there are several challenges related to data sampling, inconsistencies between instruments, and the different variability behaviour exhibited in the X-ray and UV (e.g. not all features seen in one band are necessarily manifested in the other). In general, the UV/X-ray mechanism is more complicated than originally thought and more work with high quality, well--sampled data is required.

One of the best studied AGN with \xmm\ is the narrow-line Seyfert~1 galaxy (NLS1), 1H~0707--495. Noted for its extreme X-ray variability (e.g. Gallo \et 2004) and complex spectra (e.g. Boller \et 2002; Fabian \et 2002, 2004, 2012), 1H~0707--495 was the subject of two very deep $500$~ks pointed observations in 2008 and 2010. The 2008 observations resulted in the discovery of relativistically blurred Fe L$\alpha$ (Fabian \et 2009) and a reverberation lag of 30 s and 200 s between the primary power law and blurred reflection component (Fabian \et 2009; Zoghbi \et 2010).

The Optical Monitor (OM; Mason \et 2000) functioned during both 2008 and 2010 observation epochs providing optical/UV simultaneously with X-ray data. These multi wavelength data have been used to study variability and the UV to X-ray spectral energy distribution in 1H~0707--495 (e.g Gallo 2006; Vasudevan \et 2011). The 2008 and 2010 deep observations are of particular interest for examination of correlated short-term UV/X-ray variability since the long observations of 1H~0707--495 were gathered predominantly with only one OM filter creating a uninterrupted UV light curve. For this reason, these data are the subject of this study. In this work we examine the 2008 and 2010 deep observations of 1H~0707--495 for correlated UV and X-ray variability. In the following section we describe the observations and generated light curves. In Section 3 we describe the DCF method employed in this work and present our results. We discuss our results in Section 4 and provide concluding remarks in Section 5.

\section{Observations, Data Reduction and Light Curves}

1H~0707--495 has been observed on fifteen occasions with \xmm. In this analysis we are primarily interested with two sets of four consecutive observations conducted between January 29 -- February 6, 2008 and September 13 -- September 21, 2010. During the 2008 epoch the OM operated in imaging mode with the UVW1 (291~nm effective wavelength) filter in place for the entire duration. The 2010 epoch operated with the OM cycling through seven optical to UV filters, however only data from the UVW1 filter is used. Data from these two epochs represent the longest observations of 1H~0707--495 in which UV and X-ray emission were observed (Table \ref{tab:obslog}).

The \xmm\ observation data files (ODFs) from each observation were processed to produce calibrated event lists using the \xmm\ Science Analysis System ({\tt SAS v13.5.0}). X-ray light curves were extracted from these event lists to search for periods of high background flaring and such periods have been neglected. Pile-up was negligible during the observations. The background photons were extracted from an off-source region on the same CCD close to the source. Single and double events were selected for the pn detector. For simplicity only the pn data are used here. The MOS data were determined to be completely consistent with the pn data by Zoghbi \et (2010). X-ray data were divided into three energy bands: a low-energy (i.e. ``black body'') band ($0.3 - 0.5 \keV$; XRB), a soft band ($0.5 - 1.0 \keV$; XRS) and an intermediate X-ray band ($1.0 - 4.0 \keV$; XRI). The band selections correspond with the regions of interest identified by the spectral analysis of Zoghbi \et (2010), where the intermediate band (XRI) is dominated by the direct power law component and the soft band (XRS) is reflection dominated. The lowest energy band (XRB) is thought to have significant contribution from the accretion disc and it would be interesting to examine its relation with the UV variability.

UV light curves are generated from the {\tt omichain} pipeline using {\tt SAS v13.5.0}. Each UV exposure in the 2008 epoch had an exposure time of 1200~s that is set by the ODFs with a total of 297 exposures. The UV count rate and count rate error are extracted from the individual OM exposures and the time associated to each reading is assigned as the midpoint time for the exposure. Although exposure times are set at 1200~s the actual time between UV data points are $\sim 1500$~s because each exposure was separated by a varying amount of time, usually $\sim 300$ s. UV images in the 2010 epoch had exposure times of $4400$ s, with time between exposures varying. The exposure times and delays are intrinsic to the UV data collection process set by the ODFs. The source extraction region was much larger than the host galaxy in the UV.  The host galaxy contamination was not accounted for, although its contribution is likely small (e.g. Leighly \& Moore 2004; Vasudevan \et 2011). In any event, the true amplitude of the UV variations from the AGN may appear smaller than it is intrinsically. 

A second faint UV source located on the same CCD but outside of the 1H~0707--495 extraction region was found and a light curve was extracted in the manor described above. To probe for any systemic effects a Pearson correlation coefficient was calculated between 1H~0707--495 and the secondary source. Both UV light curves were found to be completely uncorrelated ($\rho = -0.04$).

To quantify the AGN variability in each light curve we calculate the fractional variability ($F_{var}$; Markowitz et al. 2003) and the reduced, $\chi^{2}_{r}$, to a linear model (Table \ref{tab:reduced_chi2}). The amplitude of the UV variations are notably less than in the X-rays, nonetheless they are significant. The two epochs were examined separately for correlations between the light curves and found to be consistent.  The analysis presented here is for the combined 2008 and 2010 data sets.
X-ray light curves were generated in each band and were binned in time to match the exposure length of the respective UV light curves. By construction the X-ray light curves were binned uniformly with data points every $1200$ s in the 2008 epoch and $4400$ s in the 2010 epoch.  The UV light curves are unevenly sampled with a non-uniform separation in time which is set by the OM exposures.

Vasudevan \et (2011) fitted the UV to X-ray spectral energy distribution for 1H~0707--495 with a simple multicolour blackbody accretion disc and a broken power law model.  The maximum disc temperature estimated from the SED fit was between $45-50$ eV.   Assuming a geometrically thin, optically thick accretion disc the emission originates from approximately $20\ R_{\mathrm{S}}$  ($1R_{\mathrm{S}} = 2GM/c^2$) if the AGN is accreting at the Eddington rate.  

The disc emission associated with the UVW1 filter is significantly redward of the peak disc emission and the flux density is less than about $1/50$ the flux density at the peak  (Vasudevan \et 2011). We can estimate the expected time delay by considering the distance at which the UVW1 radiation is emitted from in the disc. Assuming a geometrically thin, optically thick accretion disc with no additional sources of heating (e.g. X-ray illumination) and using Wien's Law to estimate temperature, the distance is:

\begin{equation}
\frac{r}{R_{\mathrm{S}}} = \left[ \left( \frac{\dot{M}}{\dot{M}_{E}}\right)^{-1/4} \frac{M_{\mathrm{bh}}^{1/4}}{6.3\times10^{7}} \frac{b}{\lambda}\right]^{-4/3}
\end{equation}

\noindent where $\dot{M}/\dot{M_{E}}$ is the Eddington mass accretion rate, $M_{\mathrm{bh}}$ is black hole mass in units of solar mass and $b = 0.0029$ (Peterson 1997).  If we consider a range of $\dot{M}/\dot{M_{E}} = 0.01 - 1$, where higher values are likely more representative of NLS1s like 1H~0707--495, and adopt a  black hole mass of $M_{\mathrm{bh}} = 2\times10^6 M_{\odot}$ (Tanaka \et 2004), the UVW1 radiation is emitted from a radius of $r \approx 200 - 900\ R_{\mathrm{S}}$ from the black hole, which is $\sim10-50\times$ farther than where the peak disc emission originates from.  The scale corresponds to a light travel time that is $\sim 0.05 - 0.23$~d from the central black hole.   

Distances could be larger if, for example,  X-ray heating were important.  We also note that the small amplitude of the UVW1 variations ($\sim 1$ per cent, Table~\ref{tab:reduced_chi2}) indicate that only a small fraction of the entire UVW1 emitting region need be producing the fluctuations.  In any event, light travel time effects of this order should be measurable given the timing resolution and long duration of our light curves. 

\begin{table}
\caption{Sampling and variability characteristics of UV and X-ray light curves.  The epoch and energy band are given in Column 1 and 2, respectively.  Column 3 and 4 describe the number of data points and the fractional variability in each light curve, respectively.  The final column is the reduced $\chi^2$ of a linear fit to the light curve.}
\begin{tabular}{l|l|c|c|c}
Epoch & Light curve & \textit{N} & $F_{var}$ & $\chi^{2}_{r}$\\
\hline
2008 & UV (UVW1) & 297 & 0.014 & 17.9\\
 & XRB (0.3 - 0.5 keV) & 349 & 0.852 & 937.0\\
 & XRS (0.5 - 1.0 keV) & 350 & 0.860 & 1360.3\\
 & XRI (1.0 - 4.0 keV) & 349 & 0.911 & 432.7\\
2010 & UV (UVW1) & 41 & 0.008 & 30.1\\
 & XRB (0.3 - 0.5 keV) & 109 & 0.427 & 1698.8\\
 & XRS (0.5 - 1.0 keV) & 108 & 0.492 & 2794.9\\
 & XRI (1.0 - 4.0 keV) & 109 & 0.621 & 798.0\\
\hline
\end{tabular}
\label{tab:reduced_chi2}
\end{table}%

\section{Cross-Correlation Analysis}

\subsection{Discrete Correlation Function}

The typical method to determine time series correlation is the cross correlation function (CCF). However, the CCF encounters problems when applied to unevenly sampled data, measurements with errors, or time series of different lengths (Alexander 1997). There are several robust methods that can correlate time series which can accommodate the conditions described above including the interpolated cross correlation function (ICCF) (Gaskell \et 1987), the discrete correlation function (DCF) (Edelson \& Krolik 1988), and the z-transformed discrete correlation function (ZDCF) (Alexander 1997).

The DCF is a commonly used tool to probe for correlation in two time series that are unevenly sampled and for which the measurement errors are known. The DCF was developed by Edelson \& Krolik (1988) to probe inter--band UV correlation in Akn 120 and NGC 4151. Since the inception of the DCF it has been used numerous times to probe for correlations between the X-ray and UV mechanisms in AGN, for example Ar\`evalo et al. (2005), SV07, Marshall et al. (2008) and Gil-Merino et al. (2011).

The DCF is described fully by Edelson \& Krolik (1988). The first step in calculating the DCF is to identify data pairs ($X_{i}$, $Y_{j}$) from each light curve that fall within the lag bin defined by $(\tau - \delta \tau/2) \leq \Delta t_{ji} < (\tau + \delta \tau/2)$, with $\Delta t_{ji} = t_{j} - t_{i}$ where $\tau$ is the transform in time (lag) and $\delta \tau$ is the bin width. The next step is to create a set of unbinned discrete correlation coefficients ($UDCF_{ij}$), for each data pair ($X_{i}$, $Y_{j}$), 

\begin{equation}
UDCF_{ij} = \frac{(X_{i} - \bar{X})(Y_{j} - \bar{Y})}{\sqrt{(s_{X} - \bar{\sigma}^{2}_{X})(s_{Y} - \bar{\sigma}^{2}_{Y})}},
\end{equation}

\noindent where $s_{A}$ is the variance and $\bar{\sigma}^{2}_{A}$ is the mean measurement uncertainty for the data sets and whose separation in time satisfies the condition above. Given M data pairs that fall within $\Delta t_{ji}$, the final step is to sum the unbinned discrete correlation coefficients and average over the number of data pairs,

\begin{equation}
DCF(\tau) = \frac{1}{M} \sum UDCF_{ij}.
\end{equation}

\noindent An error on the discrete correlation coefficient may be calculated for each lag,

\begin{equation}
\sigma_{DCF}(\tau) = \frac{1}{M-1} \sqrt{\sum (UDCF_{ij} - DCF(\tau))^{2}}.
\end{equation}

\noindent Only data that contribute to a given lag are used in the calculation of mean and variance. Edelson \& Krolik (1988) acknowledge that binning the lag component does interpolate the correlation function but should not be confused with interpolation of data. With that in mind consideration should be made to the lag bin width, $\delta \tau$, which is a compromise between accuracy in the statistical measurement of mean, uncertainty and variance in Equation (2) and resolution of the DCF, where the former demands larger $\delta \tau$ the latter requires small values of $\delta \tau$. While there is no standard rule for choosing lag bin width, $\delta \tau$, Edelson \& Krolik (1988) argue that results depend weakly on bin width. A minimum value of $\delta \tau$ is dictated by the resolution of the coarsest time series.

\subsection{Monte Carlo Confidence Intervals}

In order to determine the confidence of the DCF we generate $10^5$ random X-ray light curves using the method described in Timmer \& K\"{o}nig (1995). We then apply the DCF with the real UV light curve and artificial X-ray light curves. The randomly generated X-ray light curves described in Timmer \& K\"{o}nig require a power law slope, $\beta$, which relate power and frequency, $S(\omega) \sim \omega^{-\beta}$. Zogbhi \et (2010) finds a power law slope of $\beta \sim 0.77$ to central frequencies which we use in this work for the three X-ray energy bands. 

Each artificial X-ray light curve is constructed without orbital gaps for an eight day period to match the total observed length in each epoch. For each Monte Carlo simulation two artificial light curves are generated, one for each epoch and each has a sampling rate to match the observed light curves in the respective epochs. Lastly, gaps are introduced into the artificial X-ray light curves by removing simulated data points that are approximately the same length as the commensurate gaps in the observed light curve.

The Monte Carlo simulations provide a distribution of $10^5$ DCF coefficients at each lag interval. The DCF coefficients were not assumed to have a Gaussian distribution so the mean and confidence intervals were not calculated using an arithmetic mean and standard deviation. Instead we construct a cumulative distribution of the DCF coefficients and find the most probable, 95 per cent and 99.9 per cent coefficients which correspond to two sided p-values of 0.05 and 0.001 respectively (Figure \ref{fig:dcf}).

We note the difficulty in obtaining the slope of the power spectrum  from the current data in either epoch given the uneven sampling. This work uses $\beta \sim 0.77$ of Zoghbi \et (2010) but the power law slope was not well constrained. The confidence intervals are sensitive to this estimate of slope and tend to be proportional to this value. For example, power law slope of $\beta \sim 1.0$ produce confidence intervals larger than those displayed in Figure \ref{fig:dcf}.

\subsection{Results}

\begin{figure*}
\includegraphics[width=1.0\textwidth]{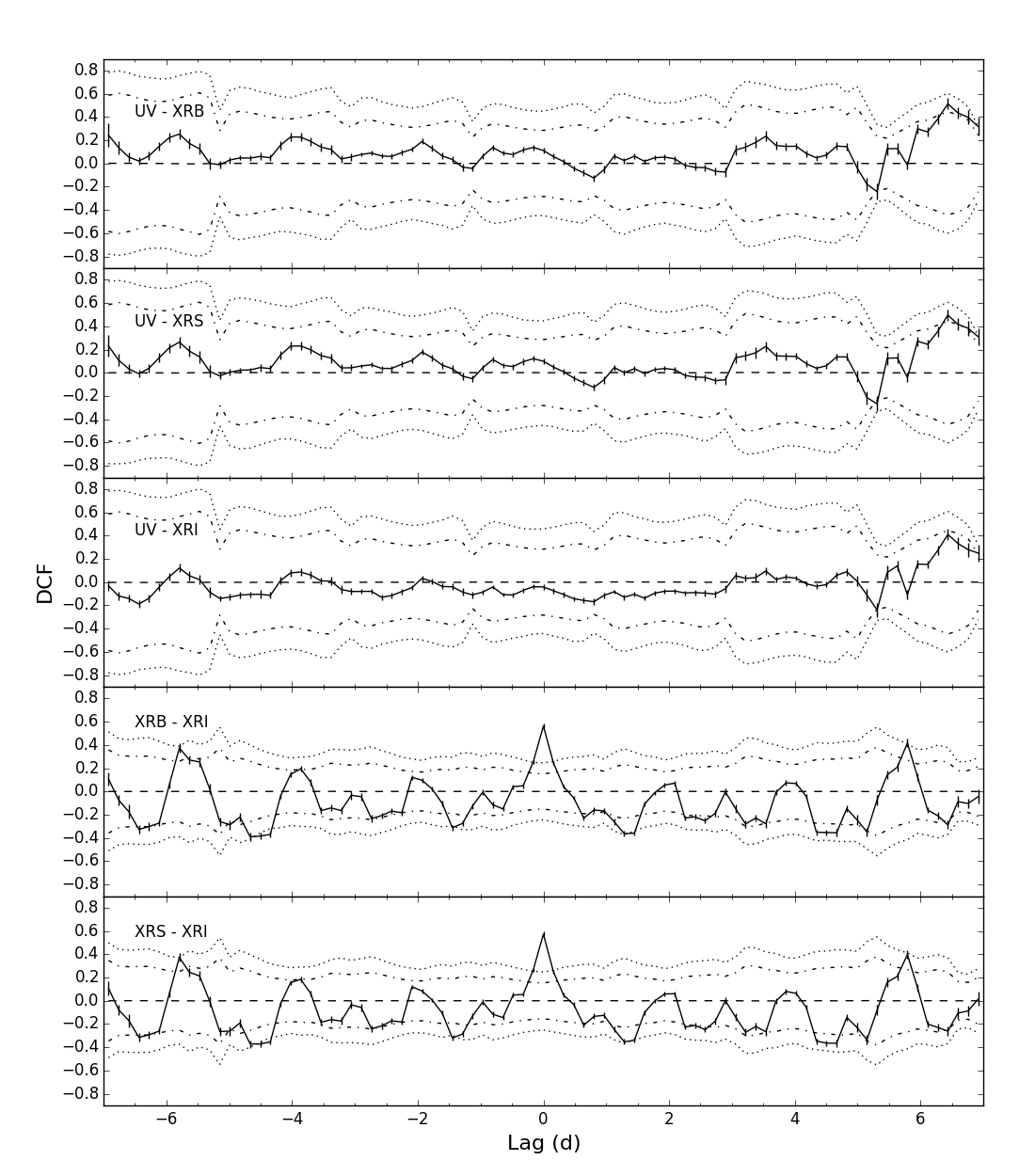}
\caption{The DCF results of combined epoch light curves. The lags are relative to the first light curve listed in the plot (negative lags indicate the UV follow the X-rays). In all plots the solid line indicates discrete correlation coefficients, the dashed line shows most probable value from Monte Carlo confidence tests while the dash-dotted and dotted lines indicated 95 per cent and 99.9 per cent confidence intervals respectively. Compared light curves are listed in each panel.}
\label{fig:dcf}
\end{figure*}

Four observed data pairs were tested with both epochs contributing to the combined light curves. The UV light curve was compared to the light curves in all three X-ray bands (UV--XRB, UV--XRS and UV--XRI) (Figure \ref{fig:dcf} top three panels). In addition, the black body and soft X-ray light curves were compared with the intermediate X-ray light curve (XRB--XRI, XRS--XRI) (Figure \ref{fig:dcf}, bottom two panel). Each DCF test used a lag bin width of $\delta \tau \sim 0.16$ d, and probed lags out to $\pm 7$~d. Lag bin width, $\delta \tau$, is a compromise between resolution and statistical accuracy. A hard minimum limit of $\delta \tau$ is set by the time resolution of the light curves, $\sim 0.05$~d.

The asymmetries in the 95 per cent and 99.9 per cent confidence intervals in Figure \ref{fig:dcf} result from the uneven sampling rate, length and gaps in the light curves. This causes varying numbers of data pairs to be found at each lag which reflect in the undulating shape of the confidence interval.

The UV--XRB, UV--XRS and UV--XRI DCFs (top three panels in Figure \ref{fig:dcf}) exhibit similar features. There is a weak ($\sim 95$ per cent confidence) anti--correlation found at $\sim 5.3$~d where the UV leads the X-ray bands. Additionally, there is a weak ($\sim 95$ per cent confidence) correlation found at $\sim 6.4$~d again with UV leading the X-ray bands. The two bottom panel in Figure \ref{fig:dcf} shows a very strong peak at zero lag for the black body (XRB) and intermediate X-ray bands (XRI) ($>99.9$ per cent) and soft (XRS) and intermediate X-ray bands (XRI) ($>99.9$ per cent).

\section{Discussion}

We have examined for time delays between the X-ray and UV light curves of 1H~0707--495 using the discrete correlation function. There were no notable lags detected between the UV and X-rays bands.  The most significant features in the DCF were at the $95$ per cent confidence level, consistent with the UV leading the X-ray fluctuations by $\sim 5.3$~d and $\sim 6.4$~d.   No features were detected at $\sim0.2$~d, which corresponds to the light-travel time between the two regions assuming a standard thin accretion disc.

The $\sim6.4$~d lag is seen as a correlation between the UV and X-ray bands.  Such a lag could arise from Compontisation of UV photons in the corona, but this is unlikely for 1H~0707--495.  Firstly, the accretion disc emission attributed to the UVW1 band is only about $1/50$ the value of the peak emission at $kT \approx 45-50$~eV (Vasudaven \et 2011) therefore such a signature will be very weak .  Secondly, analysis of the \feka\ emissivity profile suggests the corona in 1H~0707--495 is very compact with approximately $90$ per cent of the power law photons originating from  $< 15\ R_{\mathrm{S}}$ during these observations (Wilkins \et 2014). It is unlikely that significant Comptonisation is taking place at such a large distance.

A UV lead could also result from a propagation scenario (Lyubarski 1967), where fluctuations propagate in the disc first modulating the UV emitting region and eventually reach the X-ray emitting region at smaller radii.  This also seems unlikely for 1H~0707--495. At $\sim 200\ R_{\mathrm{S}}$, where we expect the UVW1 emission to originate from, we can estimate the inflow time scale in a standard accretion disc from equation 5 of LaMassa \et (2015).   Assuming a viscosity parameter of 0.1 and that 1H~0707--495 hosts a Kerr black hole (i.e. radiative efficiency, $\eta=0.3$) that is radiating at the Eddington luminosity (Zoghbi \et 2010) the inflow time scale would be several orders of magnitude larger than the observable time scales.

Reprocessing of X-ray photons in the accretion disc could make the disc material hotter than if heated by accretion alone.  If this emission is entirely absorbed and re-emitted it should show the variability associated with the X-ray source.  If the standard accretion disc around the black hole in 1H~0707--495 were being illuminated by an isotropically emitting source located at a height of $10\ R_{\mathrm{S}}$ above the disc, approximately 30 per cent additional flux would emerge in the UVW1 band than from the disc alone.  If the X-ray source were located at $100\ R_{\mathrm{S}}$ the additional flux in the UVW1 band would be approximately 40 per cent.  Given these values are significantly more than the $\sim 1$ per cent variations we are observing in the UVW1 band indicates the X-ray source may be illuminating the disc anisotropically, which is consistent with the interpretation of a compact  corona in 1H~0707--495.

The second feature in the DCF, of equally modest confidence to the $\sim6.4$~d lag, is an anti-correlation found between the UV and X-ray emission at a lag of approximately $5.3$~d.  We can speculate whether such an anti-correlation could arise in extreme gravitational environments like those present in 1H~07070--495. 
Given the centrally compact nature of the x-ray source  gravitational light bending is significant.   Depending on its height above the disc, the amount of radiation seen from an isotropic emitting source would look different for a distant observer and one at the disc (e.g. Miniutti \& Fabian 2004).   

Variability in the structure of the corona can also complicate the situation.  As Wilkins \et (2014) demonstrate the radial extent of the corona in 1H~0707-495 can vary by about 30 per cent.  While the work of Wilkins \et (2014) was based on flux-resolved spectra and did not constrain the variability time scales, there are indications from another NLS1, Mrk~335 , that these variations could occur over the course of days (Wilkins \& Gallo 2015).   

A second possibility that could lead to the anti-correlation is if the primary X-ray source were moving at a significant velocity (e.g. if it were the base of a jet).  In this case, beaming effects would modify the emission seen by the disc again making an intrinsically isotropic source appear anisotropic (e.g. Reynolds \& Fabian 1997; Beloborodov 1999; Gallo \et 2015).  These arguments are simply conjecture, but could provide motivation for further investigation, however given the modest significance of the feature we do not elaborate further.

The fact that we do not find any strong correlations between the UV and X-ray wavebands may not be surprising.  We are confident that component dilution (i.e. multiple physical components contributing to emission in a given band) is present in the X-ray bands and this is likely also the case for the UV band.  There is also the matter that the X-ray region is compact and light bending will significantly diminish the amount of X-ray light reaching the disc at $> 100\ R_{\mathrm{S}}$.  An isotropically emitting corona extending over a large fraction of the disc would generate more significant correlations between the UV and X-ray light curves.

The correlation between the XRS--XRI bands at zero lag are consistent with Zoghbi \et (2010) results finding a $\sim 30$~s reverberation lag and $\sim 200$~s lead between two similar bands ($0.3 - 1.0 \keV$ and $1.0 - 4.0 \keV$). We lack the resolution in the DCF to identify these lags separately in this study. 

The black body (XRB) and intermediate (XRI) X-ray bands are also significantly correlated with zero lag.  The black body component (see figure 8 in Zoghbi \et 2010) contributes most significantly in the XRB band.  If this is thermal disc emission generated in the inner few gravitational radii of the disc it could serve as the source of seed photons for Comptonisation in the corona producing the strong correlation with the power law dominated XRI band.
Alternatively, in the strong gravity regime close to the maximum spinning black hole returning radiation (Cunningham 1976) can heat the inner accretion disc.  If this radiation penetrates the disc to an optical depth greater than 1 it will then re-radiate as a black body.  This can lead to a correlation between the power law and black body emission (i.e. XRI and XRB).  Such behaviour was seen in the flux-resolved spectral analysis of a very similar NLS1 IRAS~13224--3809 (Chiang \et 2015).

\section{Conclusion}

The results of our analysis do not reveal any strong correlations between the UV and X-ray emitting regions in the NLS1 1H~0707--495  on time scales of less than a week.  This may not be surprising in the case of 1H~0707--495 given its extremely compact X-ray emitting region.  In such  cases any correlations are expected to be week.  Long duration observations with \xmm\ (and {\it Swift}) have provided the opportunity to begin examining UV and X-ray timing studies on longer timescales (e.g. weeks to months), but these observations require massive time investment.  Dedicated multi--band timing missions such as {\it Astro-Sat} will improve such investigations.

\section*{Acknowledgements}

We would like to thank Michael Gruberbauer who provided expertise on signal processing.  We are grateful to Daniel Wilkins for interesting discussion and help with Mathematica.  Many thanks to the referee for providing stimulating comments that have improved the manuscript.  DR is partially supported by Natural Sciences \& Engineering Research Council (NSERC) Canada USRA and PGSM award. The \xmm\ project is an ESA Science Mission with instruments and contributions directly funded by ESA Member States and the USA (NASA).

\bsp

\label{lastpage}

\end{document}